# Integrating Agent-based Programming with Elementary Science: The Role of Sociomathematical Norms

**Amanda DICKES[a,c], Amy FARRIS[a,c] & Pratim SENGUPTA[b,c*]**
[a]*Learning Sciences Program, Vanderbilt University – Peabody College, USA*
[b]*Educational Studies in Learning Sciences, University of Calgary, Canada*
[c]*Mind, Matter & Media Lab, University of Calgary, Canada*
*pratim.sengupta@ucalgary.ca

**Abstract:** How can elementary grade teachers integrate programming and computational thinking with the science curriculum? To answer this question, we present results from a long-term, design-based, microgenetic study where 1) agent-based programming using ViMAP was integrated with existing elementary science curricula and 2) lessons were taught by the classroom teacher. We present an investigation of the co-development of children's computational thinking and scientific modeling and show that the integration of programming with scientific modeling can be supported by the development of sociomathematical norms for designing "mathematically good" computational models.

## 1. Introduction

Modeling is the language of science (Lehrer, 2009; NGSS, 2013). The integration of modeling, and other epistemic practices has been recognized as a central objective for K-12 science education (NGSS, 2013). Over the past few years, the integration of computational modeling in K-12 classrooms has become an important focus of research (Sengupta, Kinnebrew, Basu, Biswas & Clark, 2013; Wilensky, Brady & Horn, 2014; Dickes, Sengupta, Farris & Basu, 2016).

The particular form (genre) of computational programming and modeling we focus on in this paper is agent-based modeling and programming (ABM). The term "agent" in ABM indicates an individual computational object or actor (e.g., a Logo Turtle), which carries out actions based on simple rules that are body-syntonic and therefore intuitive for learners (e.g., moving forward, changing directions, speeding up, etc.). Consequently, it is no surprise that researchers have been arguing for teaching and learning motion in elementary classrooms using agent-based programming since the 1980s (Papert, 1980). Many contemporary ABM platforms employ visual programming interfaces, which makes it even easier for learners to assigned or control these rules (Sengupta et al., 2015). In the context of learning kinematics, previous research shows that given appropriate teacher-led scaffolding, middle and high school students can effectively use Logo-based platforms to develop deep understandings and mathematical representations of motion (diSessa, Hammer, Sherin, & Kolpakowski, 1991; Sherin, diSessa, & Hammer, 1993; Sengupta & Farris, 2014). However, the same literature also highlights challenges in the classroom adoption of such pedagogical approaches. The high overhead associated with teaching Logo programming and teaching physics can lead the demands on the teacher to be potentially "prohibitive" (Sherin, diSessa, & Hammer, 1993, p. 116). A central challenge stems from the sequestered nature of teaching and learning programming on one hand, and teaching and learning physics using programming on the other, typically requiring a different teacher for each part (Sherin, diSessa, & Hammer, 1993). Our goal is to address this challenge by integrating, and not sequestering these two forms of instruction.

In this paper, we advance an argument that emphasizing *mathematizing* and *measurement* as key forms of learning activities, through the development of *sociomathematical norms* (McClain & Cobb, 2001; Yackel & Cobb, 1996; Cobb, Wood, Yackel, & McNeal, 1992) can help teachers meaningfully integrate programming as the "language" of science. We report a study in which a third grade teacher, in partnership with researchers, integrated agent-based programming with her regular science curriculum iteratively developing a particular form of such normative modeling practices: *sociomathematical norms* (McClain & Cobb, 2001).

## 2. Theoretical Framework: Socimathematical Norms for Integrating Programming with K-12 Science

Our previous work has demonstrated that bringing about such integration requires careful attention to the design of programming languages, as well as activity systems. Along the first dimension, we argue programming languages should employ both domain-specific and domain-general programming commands (Sengupta, & Farris, 2012; Sengupta et al., 2013; Farris & Sengupta, 2014). Along the second dimension, the design of learning activities should seek to tightly couple programming and science. For example, Sengupta & Farris (2012) and Sengupta et al. (2013) proposed an activity sequence in which initial activities can foster necessary competencies such as thinking like an agent through embodied modeling, which can also help children become proficient with programming syntax, commands and control flow, and practices such as debugging, through activities such as "drawing" simple geometric shapes with their bodies and then modeling the shapes using programming. In later activities, children can use these shapes as models of motion. These studies have shown that as students progress through these activities, they begin to become more fluent in modeling motion as a process of continuous change, which has been shown to be a key conceptual challenge for K16 students (Dykstra & Sweet, 2009).

However, research on integrating programing with the K-12 science curriculum has been largely interventionist in nature (diSessa et al., 1991; Sengupta et al., 2013; Wilkerson-Jerde, Wagh & Wilensky, 2015). In contrast, our work here takes an *integrative* stance, where our role as researchers were largely limited to designing activities in partnership with the teacher, based on what the teacher wanted to accomplish on a day to day basis, as mandated by the state and national science and math standards. We believe that such forms of partnership are methodologically crucial for addressing the issue of effectively managing the tradeoff between teaching programming and teaching science.

In this paper, we propose that emphasizing mathematizing and measurement as key forms of learning activities can help teachers meaningfully integrate programming as a "language" of science, and further, that teachers can accomplish this by supporting the development of sociomathematical norms. The iterative design of mathematical measures can result in deep conceptual growth of students in elementary science, especially when these activities are integrated throughout the curriculum over several months (Lehrer, 2009). Furthermore, the development of children's scientific and mathematical modeling in the classroom in an authentic manner should also involve and can be greatly benefitted by the iterative development and refinement of collective, (i.e., classroom-level), *normative* modeling practices (McClain & Cobb, 2001; Lehrer, Schauble & Lucas, 2008). Sociomathematical norms (McClain & Cobb, 2001; Yackel & Cobb, 1996; Cobb, Wood, Yackel, & McNeal, 1992) differ from general social norms that constitute the classroom participation structure in that they concern the normative aspects of classroom actions and interactions that are specifically mathematical. These norms regulate classroom discourse and influence the learning opportunities that arise for both the students and the teacher.

An important, and rather fundamental sociomathemtical norm is what counts as an *acceptable* mathematical solution, and further, this norm typically originates as a socially defined norm, and shifts over time to a more sociomathematically defined norm (Yackel & Cobb, 1996). Similarly, science educators have also shown that the question of what counts as a "good" model also needs to be normatively established in classroom instruction in order to deepen students' engagement with scientific modeling in elementary grades, and that these norms also follow similar shifts toward deeper disciplinary warrants over time (Lehrer, Lucas & Schauble, 2008). Our goal is to demonstrate how the emphasis on developing and refining sociomathematical norms pertaining to the design of mathematical measures of motion can help teachers seamlessly integrate programming with science education in a 3$^{rd}$ grade classroom.

## 3. Research Questions

Specifically, we investigate the following research questions:

1. What were the sociomathematical norms that developed, and how were they taken up by the students?
2. Did these norms shape in any way the development of students' computational models and computational thinking? If so, how?

## 4. Method

### *4.1 The Programming Environment*

We used ViMAP (Sengupta, Dickes, Farris, Karan, Martin & Wright, 2015), an agent- based, visual programming language that uses the NetLogo modeling platform as its simulation engine (Wilensky, 1999). In ViMAP (Figure 1), users construct programs using a drag-and-drop interface to control the behaviors of one or more computational agents. ViMAP programming primitives include domain-specific and domain- general commands as well as a "grapher" which allows users to design mathematical measures based on periodic measurements of agent-specific and aggregate-level variables (e.g., speed and number of agents, etc.)

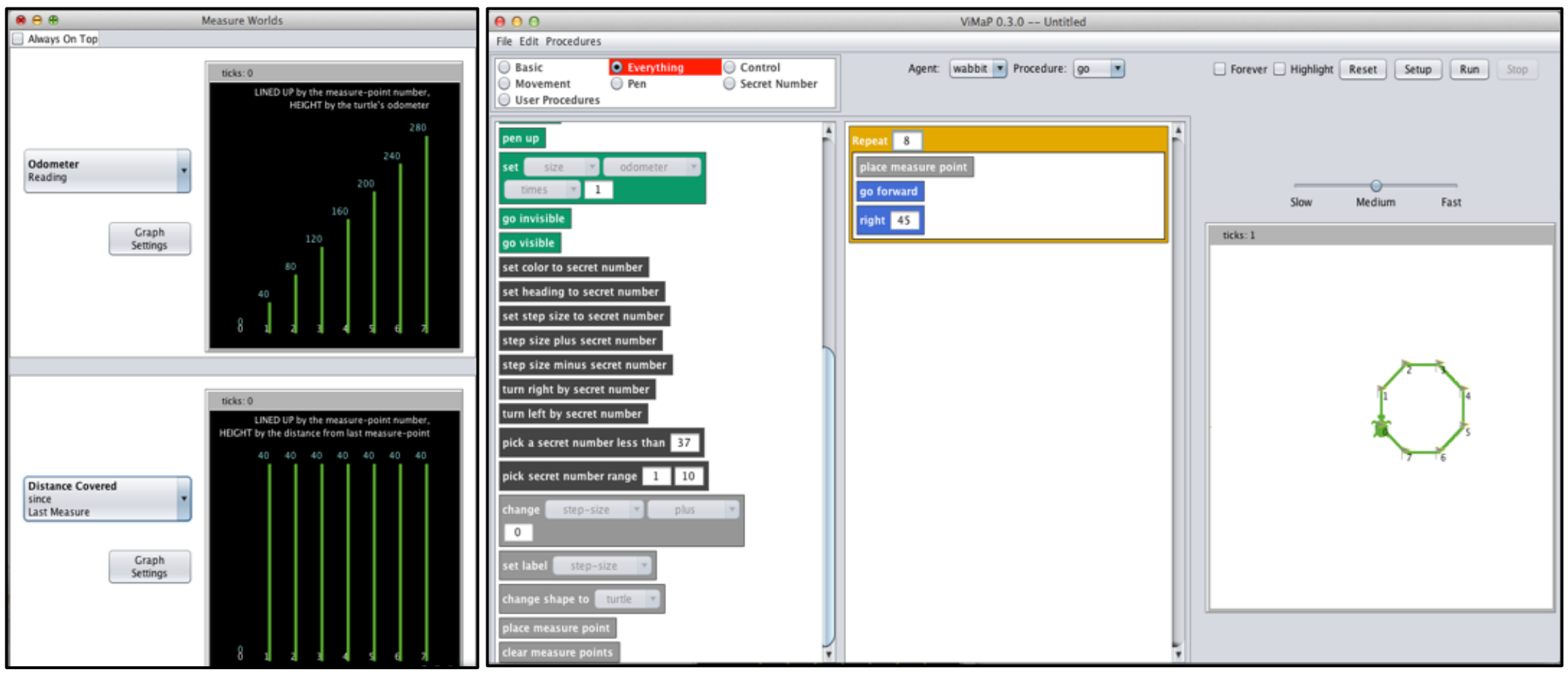

<u>Figure 1</u>: ViMAP's measurement window and programming interface. Figure illustrates the program for generating a regular octagon, the enactment by the turtle agent and graphical representations of length of each line segment (graph on lower left) and perimeter (graph on top left)

### *4.2 Setting & Participants*

<u>Table 1: Summary of learning activities during Phase II</u>

| Activity | Description |
|---|---|
| Leaving Footprints | Students leave ink footprints on banner paper. |
| Generating Measures | Students iteratively develop, apply, test and refine a measurement of distance termed a 'step size'. |
| Collecting step-size data | Students use the 'step size' measurement convention to measure their personal step sizes. |
| Modeling Step-sizes in ViMAP | Students model their personal step-sizes in ViMAP. Total-distance graphs & predictions using ViMAP's grapher are generated and discussed. |
| Modeling Motion as a Process of Continuous Change | Students model motion scenarios in ViMAP and check the validity of those models using ViMAP's grapher and the total distance equation. |

This study was conducted over the course of 7 months in a 3rd grade classroom in a 99% African-American public charter school located in a large metropolitan school district in the southeastern United States. Fifteen students – fourteen African-American and one Latino – participated in this study. Researchers met weekly with the classroom teacher and iteratively co-designed the classroom activities. The teacher taught all lessons, and changes to the activities were made based on the her formal and informal assessments of student understanding of the material or in-the-moment responses to student ideas. These adjustments often took the form of extending instructional time on a topic, and modifying the designed classroom materials to better meet the mandated instructional goals. Throughout the year, the teacher emphasized connecting modeling in ViMAP to other out-of-computer modeling experiences, such as embodied and physical modeling activities, as well as re-framed the computational representations in ViMAP as analogous to meaningful lived experiences for both herself and the students. The emphasis on developing classroom-wide conventions was a practice that the teacher employed in her regular math instruction. Our study focuses on how the teacher adapted and employed this approach as a way to integrate modeling motion using ViMAP with her science curriculum.

The learning activities were divided into three phases: Phase I (Geometry), Phase II (Kinematics) and Phase III (Ecology). The present paper reports on Phase II, Kinematics, and traces the development of three normative, mathematical criteria for "what counts" as good ViMAP models of motion. Instruction during Phase II focused on the invention and interpretation of mathematical measures and using ViMAP as a way to explain a real-life phenomenon involving motion (e.g., walking at a constant rate or two cars traveling at different rates for different periods of time). Table 1 summarizes the learning activities during Phase II.

### *4.3 Data & Analysis*

Data for this work comes from informal interviews with the participants, video recordings of class activities and discussion, student artifacts (e.g. student representations, activity sheets, ViMAP models and pre-, mid, and post-assessments) and daily field notes. The lead researcher and the classroom teacher conducted informal interviews during opportune moments while the students were engaged in single, pair or small group work around modeling and representational activities. Classes were video recorded, and student-created artifacts (ViMAP models, written work) were also collected.

We present the analysis of in the form of explanatory case studies (Yin, 1994), which are well suited as a methodology to answer *how* and *why* questions. We find this to be good fit because our goal here is to illustrate the *process* through which the classroom developed sociomathematical norms, which includes answering *how* the development of these norms shaped the students' interactions with ViMAP and other modeling experiences, and *why* these norms were deemed useful by the teacher. Following previous studies (Dickes et al., 2015), our selection and analysis of cases were guided by the following two criteria: representativeness and typicality.

Representativeness implies that the selected cases should aptly represent key aspects of learning experienced by the students. These key aspects or themes, in turn, are defined based on the research questions. For our purposes, representativeness implies that each case should highlight an important aspect of the process through which the relevant sociomathematical norm emerged. *Typicality* implies that the selected case(s) should potentially offer insights that are likely to have wider relevance for the remainder of the participants in the study. In other words, the cases selected should represent aspects of the process of learning experienced by a majority of the student population. Each of the cases we presented here was typical of majority of the students in the classroom (>80%), as evident in comparisons of student work across all students. In addition, to answer RQ2, we also present a classroom-level analysis of students' ViMAP code and models in terms of the quality of their code. We explain the coding scheme later, along with the Findings, and also explain why we believe these changes in students' computational work were shaped by their take-up of the sociomathematical norms.

## 5. Findings

The analysis presented below illustrates the development of sociomathematical norms for measuring (Inventing Measures), describing (Approximations) and extending data (Predictions). We explain these norms, describe how they were taken up in student work and how the development of each norm paralleled the new computational practices.

### *5.1 Inventing Measures: Movement from Social to Sociomathematical*

At the start of Phase II, students generated an embodied artifact – their own footprints inked onto a strip of banner paper. After this activity, the classroom teacher problematized the idea of a step size. What is a step size and if we were to measure one, where would we begin measuring and where would we end? Students offered three options for 'step size' measures. Step sizes are measured from heel-to-toe, measured from heel-to-heel and finally measured from toe-to-toe (Figure 2). At this stage, choosing which step size was best was primarily a social endeavor, with students defining the best step size measure based on a class vote, ultimately selecting the heel-to-toe measurement convention because it was quote, "the biggest" or because their "friend voted for it", indicating the "socially" grounded nature of what counts as a good measure.

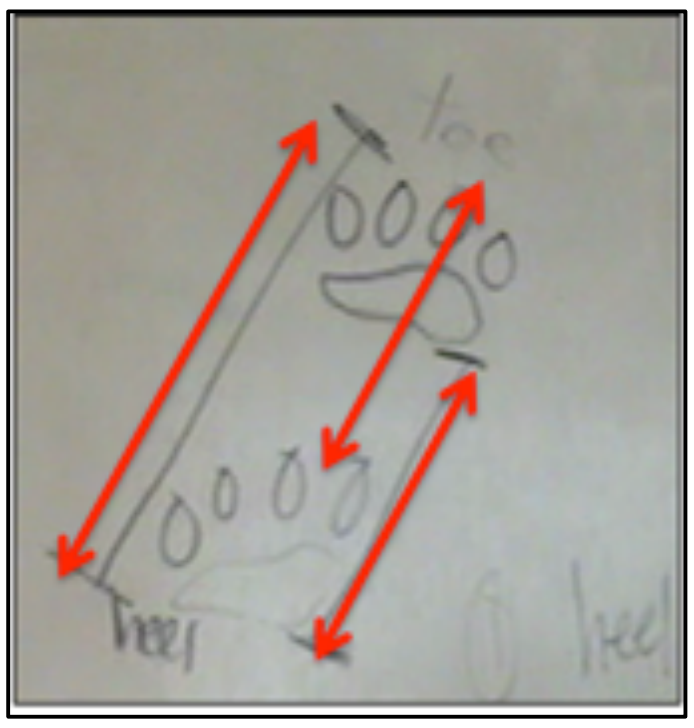

Figure 2: Student ideas on how to measure a "step size"

To problematize their selected measurement convention, the classroom teacher asked the students to return to their footprint artifacts and measure their unique step sizes using the heel-to-toe measurement convention. The classroom teacher then asked for students to add up each of those individual step sizes to generate a total distance travelled. Finally, the classroom teacher asked the class to measure, with yardsticks or measuring tape, their total distances traveled on their footprint artifacts and record that value on the same data sheet. Students reported their findings and found that the measured (ruler on footprint artifact) and the calculated (adding step sizes measured using convention) "didn't match", when they had predicted that they would (Figure 3).

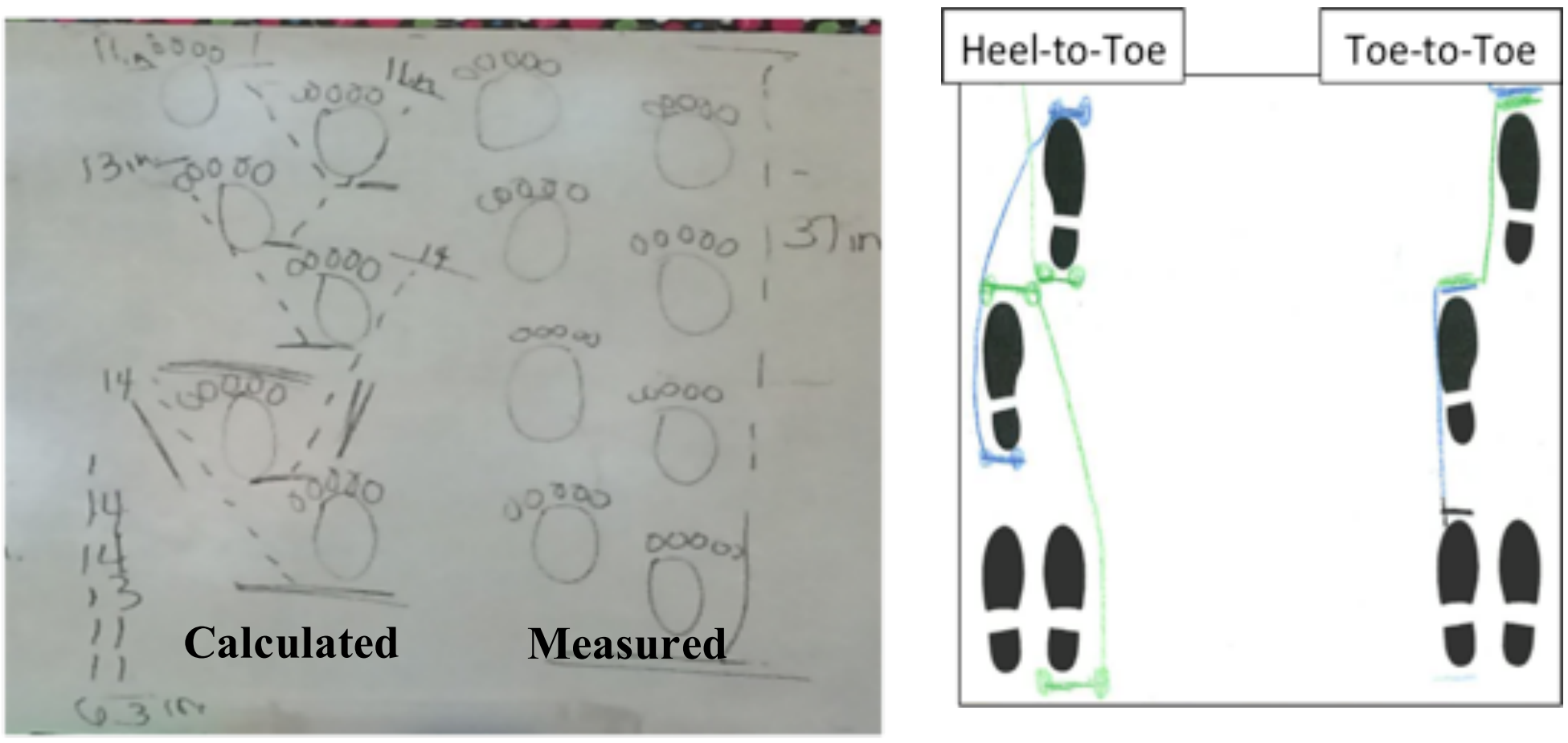

Figure 3: Refinement of "step-size" measurement convention from socially defined (heel-to-toe) to sociomathematically defined (toe-to-toe).

The measured and calculated total distances did not match *due to the selected measurement convention*. The heel-to-toe measurement convention produces an overlap, effectively measuring parts of the total distance twice (63 in. vs. 37 in. in Figure 3). This discrepancy prompted the classroom teacher to suggest to the class that "maybe we need to find a measure that is more mathematically accurate". The outcome of this disruption was the invention of the "mathematically accurate" toe-to-toe measurement convention, shown in Figure 3. What is notable in this episode is the development of criteria for what counted as "good measure". Initially, "good" measures were socially defined, with students selecting how to measure a "step-size" for reasons unrelated to the purpose of the measure. Following the failure of the heel-to-toe convention, the criteria for "good" measures shifted towards mathematical accuracy. In other words, the value of the measure was assessed on its ability to accurately measure what the measurer intended for it to measure.

*5.2 Approximation & Prediction: Norms for Model Refinement*

In addition to developing measurement conventions for step sizes, students also explored ideas of approximation and prediction as methods for summarizing and extending data and refining their ViMAP models. After students had re-measured their step sizes based on the new toe-to-toe measurement convention, the class was asked to think about what value best represented their individual step size data. The teacher introduced 'approximations' because, as she explained to the researchers, she believed that averages would be difficult for her students. She framed an approximate value by discussing a few examples from their embodied step size activity, where she explained an approximate value as a value that is "close to the actual but not exact", and at the same time, represent the general trend of the values.

Table 2: Angelo's prediction

| | Utterance | Line |
|---|---|---|
| Researcher | How far did you walk after taking 15 steps? | 1 |
| Angelo | 300 distance | 2 |
| Researcher | That's exactly right. | 3 |
| Angelo | So, if somebody bet that I won't make it farther than 100 *I know* that I will | 4 |
| | make it. | 5 |
| Researcher | That's right. That's how a formula for approximate distance can help you. | 6 |
| | If someone said "I bet Angelo would only walk 150 inches in 15 steps", but | 7 |
| | knew what your approximate step size was, could you prove them wrong? | 8 |
| Angelo | Yes | 9 |
| Researcher | How? | 10 |
| Angelo | I could look at my graph. | 11 |
| Researcher | Or you could do what? | 12 |
| Angelo | I could use a calculator. Fifteen times 20 equals 300. | 13 |

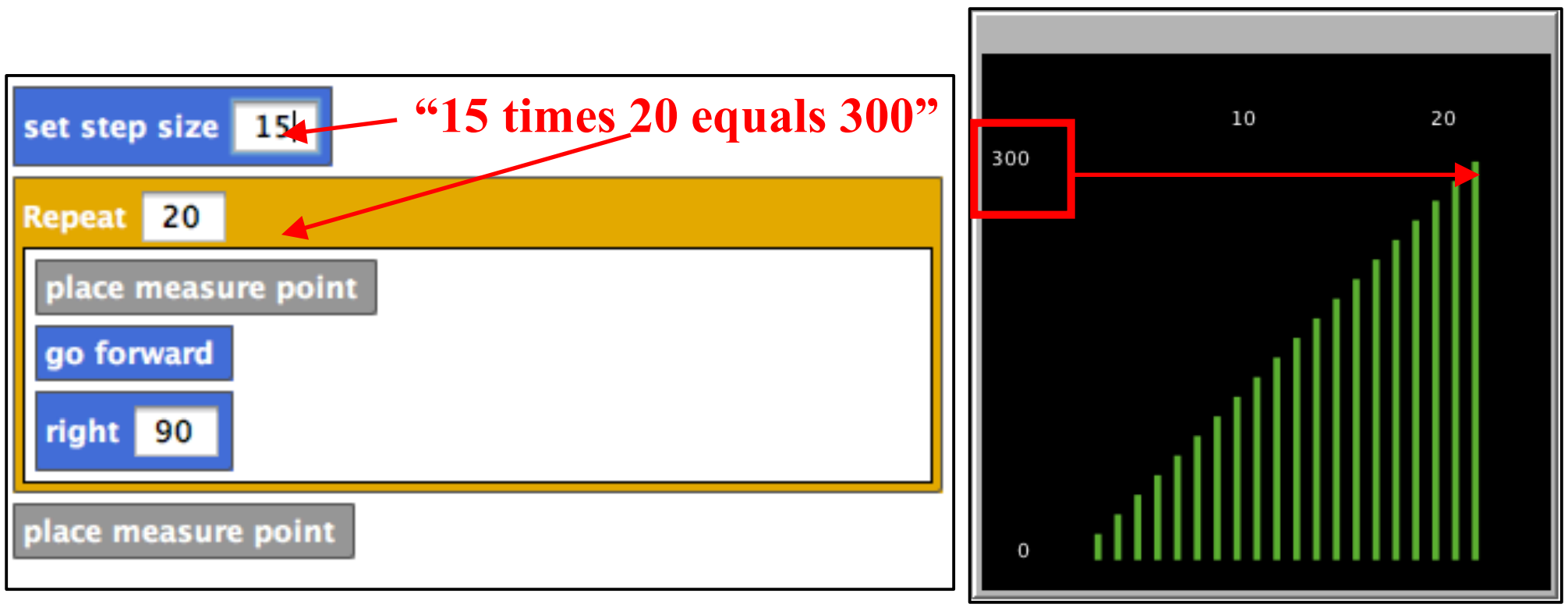

Figure 5: Angelo's ViMAP model

The normative nature further developed through and became evident in the form of class discussions where deviances from the norm were addressed. For example, the teacher provided students with a hypothetical data set, step sizes of 11, 9, 11 and 12, and asked them to build a ViMAP model of the total distance traveled based on the general pattern of the step sizes. To facilitate this, she asked students to reason about the following: if the hypothetical student "continued walking", what would "their next step be"? In a flurry of discussion, each of the fifteen students offered their ideas. Fourteen of fifteen students (93%) agreed that a "good" possible "future step" was any value already within the range of the set of empirical data, i.e. a value of 9, 10, 11, or 12. One student in particular offered to the class that "11" was the best choice because it appeared "the most times" and was "in the middle" of the data set. Only one student deviated from the other students, suggesting that "13" was the next possible step since it "continued the pattern" established by the final two steps of 11 and 12. The teacher referred back to the shared classroom definition of approximation, *close to actual but not exact,* and modeled in ViMAP an approximate step size of '13'. She then asked the class to consider the total distance travelled in each model: 43 using actual step size values and 53 using an approximate value of 13. The class agreed that the two distances were not 'close' and came to a consensus that 'good' approximate values were "close to" the actual value in terms of both individual step sizes *and* total distance traveled.

Playing with approximate values also gave students predictive power through extending their ViMAP models of motion and using multiplicative reasoning. During a teacher-led class discussion on calculating approximate total distances, students noticed that you could use repeated addition (8+8+8+8+8+8) or multiplication (8 x 6) to quickly solve for total distance traveled using approximate steps sizes. The teacher asked what "the formula" for finding total distance would be if they were not "using numbers". Two students in the class responded that they are multiplying the "number of steps" by the "approximate step size", generating a formula for total distance: *Total Distance = Number of Steps x Approximate Step Size*. In the teacher's words, this formula would allow the students to "find total distances that you can't actually walk", and therefore, can be used to make predictions.

How did students take this up in their work? An illustrative case is shown in Table 2. In this excerpt, Angelo (a student) interprets the formula as a means to both "win a bet" as well as mathematically verify the accuracy of his ViMAP model of distance. Angelo comments in lines 4 and 5 that if someone bet him that he would only travel less than or equal to 100 units of distance, he would know that they were wrong based on his understanding of approximate values and their role in the total distance formula the class had derived. The researcher affirms Angelo's observation, asking him if he could prove an acquaintance wrong if he knew his approximate step size (lines 6, 7 and 8). Angelo responds in line 8 that he could and when asked by the researcher how he could prove them wrong (line 9), he offers two possible solutions: the graphs he had generated in his ViMAP model (shown in Figure 5) and his formula (line 13).

Epistemologically, this is a significant move. As Angelo put it, using approximate values allows him to "know" (line 4). We believe that Angelo's explanation of "betting" and "knowing" here is his intuitive way of explaining what prediction is. Furthermore, this demonstrates that Angelo is able to mathematically summarize discrete values to model continuous patterns of change.

### *5.3 Further into Prediction: Generalizing Motion using a Multiplicative Scheme*

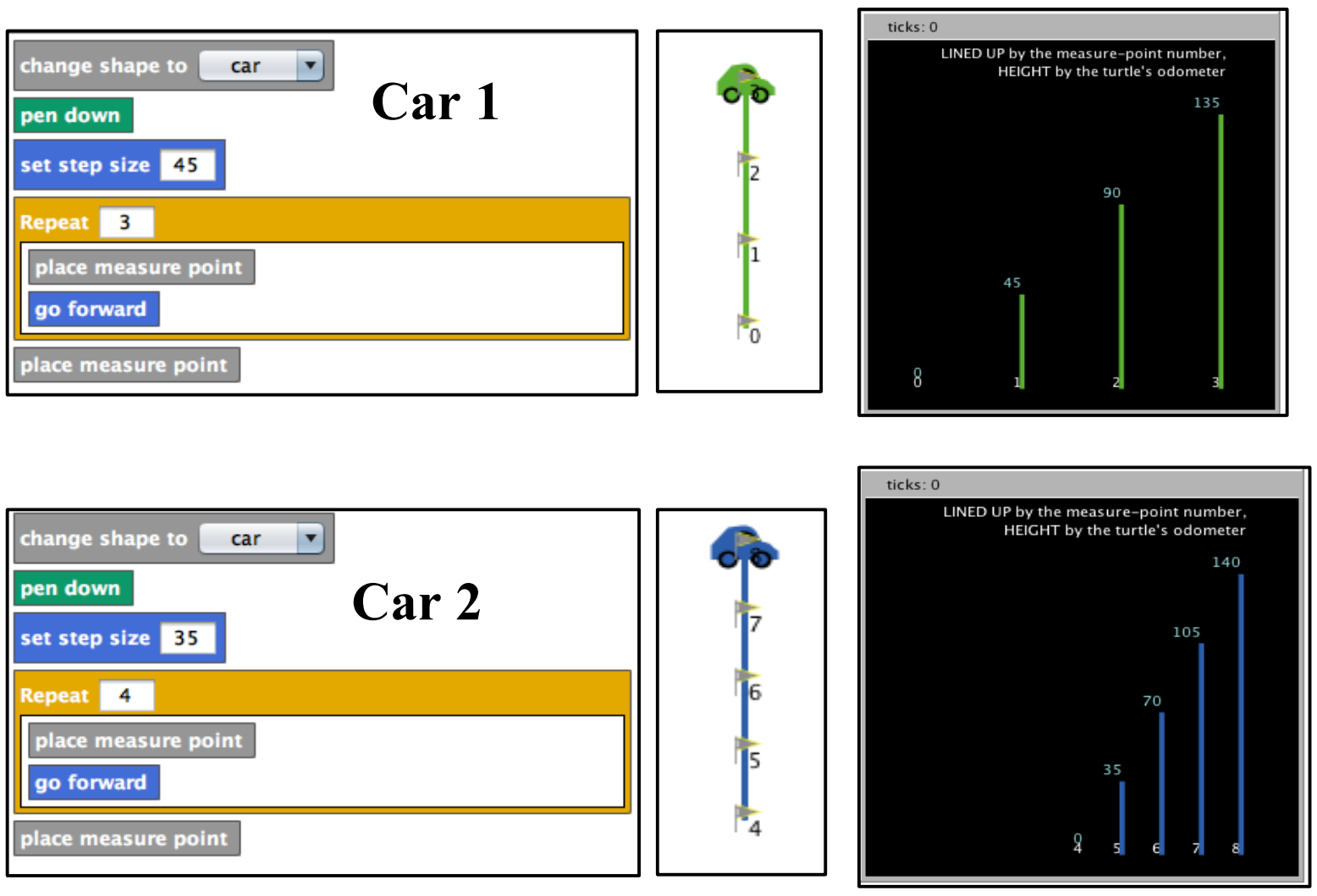

Figure 6: A student's solution to the two-car problem using ViMAP

Toward the end of Phase II, the classroom teacher and researchers wanted to extend the thinking students had done on developing predictive models of motion into more generalizable mathematical forms. The teacher recognized that the formula for Total Distance derived by the class (Number of Steps x Approximate Step Size) was a specialized form of the a multiplicative scheme that also serves as a rate equation: *Distance = Speed x Time*. She told the researchers that she considered this to be a great context for engaging her students in multiplicative reasoning. She explained to the class that this is a "powerful" formula, which can be used to analyzed many real-world situations. She then introduced a "real world" problem, in which students had to of figure out which of two cars, Car 1 or Car 2, traveled further. Car 1 traveled at a speed of 45 mph for 3 hours, while Car 2 traveled at a speed of 35 mph for 4 hours. A sample student's work is shown in Figure 6. As students shared their ViMAP models, we noticed that all of them were able to produce ViMAP models that used appropriate and non-redundant variables. The multiplicative reasoning was evident in students' use of "repeat" and "step-size", as shown in Figure 6, where Car 1 travels 3 (repeat) x 45 (step-size) units, and Car 2 travels 4 (repeat) x 35 (step-size) units.

### *5.4 Co-development of Sociomathematical Norms and Computational Thinking*

Our analysis also shows that across the class, there was also an increase in students' ability to compose ViMAP models that accurately represented their data. Students' use of, as well as their skill, at generating accurate ViMAP graphs also increased over Phase II. The growth in students' computational fluency is evident in Figure 7, which shows how students' use of the ViMAP programming commands became increasingly sophisticated as they held their models accountable to the sociomathematical norms throughout the duration of the activities reported in the paper (Phase II). We scored each student's final ViMAP model at the end of each class period in terms of whether they used appropriate variables in their ViMAP code, and whether their graphs represented appropriate element(s) of the phenomenon being simulated using their ViMAP code, each on a scale of 0 – 3. A score of zero meant none of the variables used were appropriate, whereas a score of 3 meant no use of redundant or incorrect variables. The accuracy of the graphs in students' later models were indicative of the appropriate use of the "repeat" command, and order of placement of the "place measure" command. This in turn relied on a conceptual understanding of when to initialize the measurement, and how often the desired measurement had to be repeated in order to generate the graph.

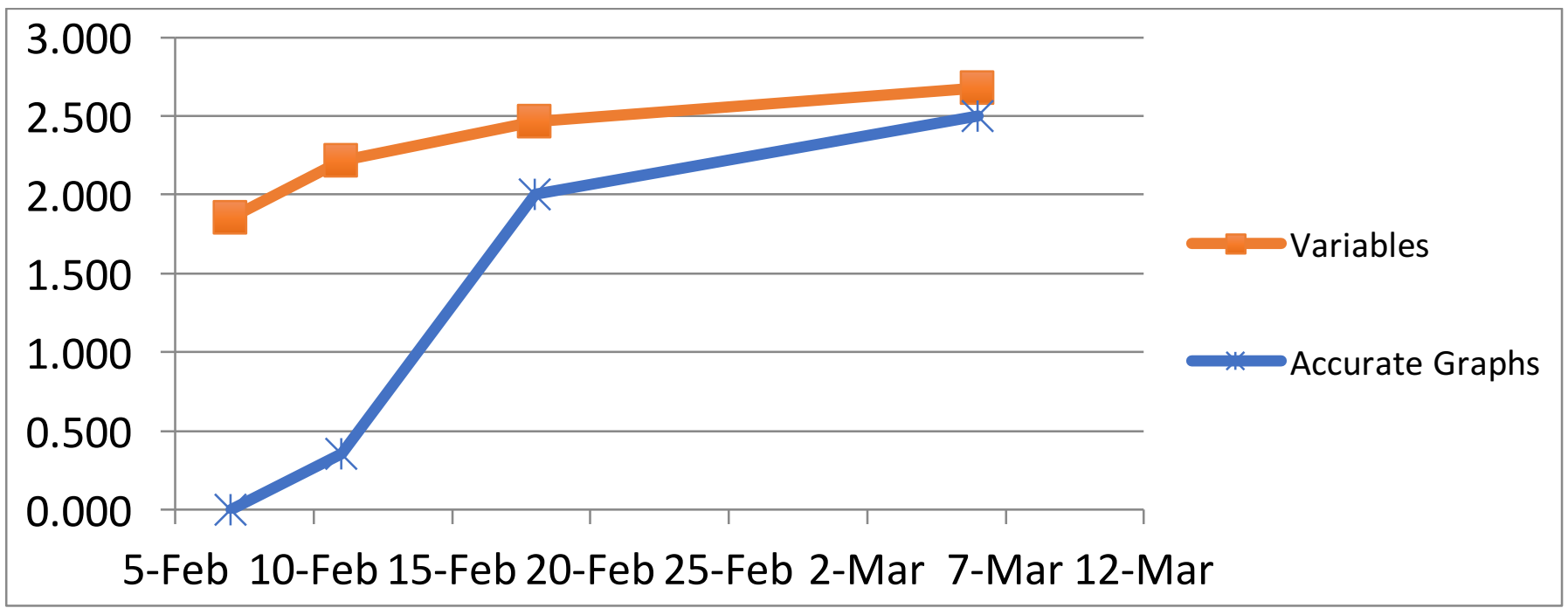

Figure 7: Improvement in Computational Thinking

Why did this improvement happen? We believe that the illustrative cases we presented shows that the development, deployment and refinement of sociomathematical norms led to iterative improvement in the quality of students' models as progressively more *authentic* representations of the phenomena they were modeling. This was reflected in the teacher's push for "accuracy", which was often taken up by students in their modeling work, and became a disposition that was "taken as shared" (Cobb, Wood, Yackel, & McNeal, 1992) by the classroom community. The push for making their models "predictive", in turn, resulted in deepening of students' multiplicative reasoning through the use of ViMAP programming, and this was evident in their use of loops and agent-level variables (No. of Repeat x Step size), as well as a more careful attention to the design of graphs. Programing in ViMAP was no longer seen as extraneous to learning science; rather, the establishment of sociomathematical norms reified the use of ViMAP programming as the *language* of doing science.

Furthermore, we also believe that the teacher's emphasis on using physical and embodied modeling as a way to complement computational modeling and thinking played an important role in the students' take-up of the norms. Cobb and colleagues have argued that sociomathematical norms pertaining to what counts as an acceptable mathematical explanation and justification typically have to be interpretable in terms of actions on mathematical objects that are experientially real to the listening students, rather than in terms of procedural instructions (Cobb, Wood, Yackel, & McNeal, 1992). In our study, by emphasizing embodied modeling as a way to mathematize motion, the teacher facilitated the students' take-up of norms pertaining to "what counts as a good model" of motion, by making ViMAP commands such as "step-size" experientially real to the students.

## 6. Discussion

### *6.1 Sociomathematical Norms Integrate Computational Thinking and Science*

Our study highlights the *reflexive* relationship between computational thinking, scientific modeling and mathematical thinking when agent-based programming is the computational medium. While this has been noted previously in researcher-led studies (Kafai & Harel, 1991; Papert, 1980; Sengupta et al., 2013), our work here shows that teachers with no background in programming can integrate programming with their existing science curricula by reframing programming as mathematization – in particular, designing measures of change. Furthermore, our study also shows that using agent-based programming as the means to develop these models of change can be supported by the teacher by developing sociomathematical norms around the mathematical quality of these models.

### *6.2 Methodological Concerns: Teacher Voice and Conceptual Dissonance in Researcher-Teacher Partnerships*

Design-based researchers have recently begun advocating for greater teacher voice and agency in research studies, which in turn reframes studies as researcher-teacher partnerships (Severance, Penuel, Sumner & Leary, 2016). Our study is certainly an example where teacher voice often led the direction of research; but it also raises an important methodological and epistemological question: how should we address conceptual dissonances between the researchers and the teachers? For example, in our study, the teacher's framing of "accuracy" - i.e., students' models must be "mathematically accurate" - was largely based on her intuitive conceptualization of the term. Let us now imagine answering this question as educational researchers and epistemologists. "Accuracy" will take on a very different meaning, and perhaps even have a negative connotation - because an essential characteristic of models, according to the epistemologists of science, is that they are incomplete. In fact, a few months later, the teacher did introduce the notion of incompleteness (albeit in her own language, and in a different context) – in Phase III, while modeling ecological interdependence. The notion of accuracy, though, lingers throughout the academic year.

We will take up this issue in more detail in a different paper. But we do want to raise the following question here: what should we do in such situations? Should we have intervened and coached the teachers about the professional vision of scientists and epistemologists about accuracy and incompleteness of models? This study is an example where we did not intervene to bridge conceptual dissonance on this issue. Our decision stems from the fact that researchers must fundamentally position teachers as the director of the partnership – rather than at an equal footing with the researcher. An equitable partnership may not be one in which everyone has equal say. Instead, an equitable partnership in educational computing research must seek to support teachers in voicing (and re-voicing) computation from their own perspectives, with curricular mandates and classroom constraints in mind.

As Heidegger famously remarked, the essence of technology is nothing technological (Heidegger, 1954). Rather, it is the "frame" in which technology lives – its lifeworld of human experience – that defines it. Unfortunately, researchers in educational computing – in particular, programing languages for children – have traditionally not engaged with the issue of curricular integration from the perspective of K-12 teachers. Research studies in this field (including some of our earlier work), therefore, largely carry out a strong interventionist agenda where teacher voice is often overshadowed by the researchers. In contrast, we have come to see the K-12 public school classroom as a complex, interdependent system, where teachers, students, curricula and curricular mandates – must all be considered alongside one another, especially if we set out to integrate any new literacy and/or technology with the classroom. So, if our goal is to make programming and computational modeling ubiquitous in the K-12 science classroom, we posit that researchers and designers of programming languages for the K-12 classrooms must learn to see the world through the eyes of the teachers, especially when it involves conceptual dissonance between researchers and teachers. It is through carefully studying the unfolding of such dissonances over longer periods of time (i.e., not a short intervention study), especially when teachers are working with new technologies and literacies (such as programming and computational modeling), that we (as researchers) will learn to design technological and activity systems that will be aligned with the perspectives of the teachers, and therefore, have a greater chance of becoming a mainstay in their classrooms.

**Acknowledgements**

Funding from the US National Science Foundation (NSF CAREER Award #1150320) and the Imperial Oil Foundation (Canada) is greatly acknowledged.